\documentclass[fleqn,10pt]{wlscirep}
\usepackage[utf8]{inputenc}
\usepackage[T1]{fontenc}
\usepackage{lineno}
\usepackage{rotating}
\usepackage{siunitx}
\usepackage{ragged2e}
\usepackage{array,booktabs,makecell,tabularx}
\usepackage{tabulary}
\usepackage{colortbl}

\title{Sub-cortical structure segmentation database for young population}

\author[1,*]{Jayanthi Sivaswamy}
\author[1,$\dag$]{ Alphin J Thottupattu}
\author[1,$\dag$]{Mythri V}
\author[1,2]{ Raghav Mehta}
\author[3]{ R Sheelakumari}
\author[3]{Chandrasekharan Kesavadas}
\affil[1]{Center for Visual Information Technology, International Institute of Information Technology (IIIT), Hyderabad, Telangana, India}
\affil[2]{Probabilistic Vision Group, Centre for Intelligent Machines, Department of electrical and Computer Engineering, McGill University, Montreal, QC, Canada}
\affil[3]{Department of Imaging Sciences and Interventional Radiology, Sree Chitra Tirunal Institute for Medical Sciences and Technology, Thiruvananthapuram, Kerala, India}

\affil[*]{corresponding author: Jayanthi Sivaswamy (jsivaswamy@iiit.ac.in)}

\affil[$\dag$]{these authors contributed equally to this work}

\begin{abstract}
Segmentation of sub-cortical structures from MRI scans is of interest in many neurological diagnosis.  Since this is a laborious task machine learning and specifically deep learning (DL) methods have become explored. The structural complexity of the brain demands a large, high quality segmentation dataset to develop good DL-based solutions for sub-cortical structure segmentation. Towards this, we are releasing a set of 114, 1.5 Tesla, T1 MRI scans with manual delineations for 14 sub-cortical structures. The scans in the dataset were acquired from healthy young (21-30 years) subjects ( 58 male and 56 female) and all the structures are manually delineated by experienced radiology experts. Segmentation experiments have been conducted with this dataset and results demonstrate that accurate results can be obtained with deep-learning methods.  Our sub-cortical structure segmentation dataset, Indian Brain Segmentation Dataset (IBSD) is made openly available at \url{https://doi.org/10.5281/zenodo.5656776}.
\end{abstract}
\begin{document} 

\flushbottom
\maketitle

\thispagestyle{empty}

\section*{Background \& Summary}
The brain is the most complex organ in the human body and imaging is critical to assess its structural and functional aspects. While a wide range of modalities are available to image the brain, the most widely used ones are Computed Tomography and Magnetic Resonance Imaging (MRI) as they aid assessment of the structural aspects. Many automated image analysis techniques have been developed to extract or segment the sub-cortical structures \cite{seg1,seg2,seg3,seg4,rag,psi}. Automating segmentation is attractive since manual delineation of the structure boundaries (by experts) is very tedious. The sub-cortical region of the brain is of key interest as it plays many vital roles such as emotion control, hormone production, memory etc. Degradation of this region results in many neurological disorders such as Alzheimer's disease \cite{Alzheimer}, Huntington's disease \cite{Huntington}, Supra-nuclear Palsy \cite{SupranuclearPalsy}, Schizophrenia \cite{Schizophrenia}, depression \cite{depression} etc. Volumetric and shape analysis of sub-cortical structures of brain is commonly used to study such disorders and hence accurate algorithms are needed for sub-cortical structure segmentation. The success of Deep learning (DL) in computer vision has also led to its application in the medical domain including for developing algorithms for medical image segmentation. DL is largely a data-driven paradigm which means data plays a critical role in training and a developing a models for various tasks including segmentation. Hence, there is a requirement for a dataset with adequate number of high quality images for a variety of tasks such as segmenting structures of interest, aligning/  registering images, etc. However, the number of 3D datasets for brain image segmentation available for public access is limited and the number of images in these datasets is also not large enough to enable building of accurate models for segmentation. 
Some of the widely used public datasets for brain segmentation and the number of MR volumes and structures with markings/labels are: i) MICCAI 2012 \cite{MICCAIdataset} with 35 MR volumes and 134 labeled structures ii) IBSR \cite{ibsr} with 20 volumes and 43 labeled structures and iii) LPBA40 \cite{lonii} with 40 volumes and 56 labeled structures iv) Hammers67n20 \cite{hammer} with 20 volumes and 67 labeled structures and v) Hammers83n30 \cite{hammer} with 30 volumes and 83 labeled structures. Besides these, there are few public structure-specific datasets with only   (ex. hippocampus, \cite{hippo}).  private datasets such as introduced by Babalola et al. provides 270 volumes and labels (via semi-automated segmentation) for 18 structures. The neuroimages in these datasets are of young to elderly individuals. 
\\
We present an Indian Brain Segmentation Dataset (IBSD), for sub-cortical brain segmentation. This has 114 MR volumes generated under a fixed imaging protocol. Each volume has 14 labeled sub-cortical structures. The number of MR scans in the dataset is of approximately equal number of male and female subjects belonging to a young age group (20-30 years). This data has been used to create a template for the young Indian population \cite{IBA100}.
Large number of high quality MR images and segmentations is the main feature of our dataset.

\section*{Methods}
\subsection*{MRI Image Collection}
\begin{figure}
\centering
\includegraphics[scale=0.7]{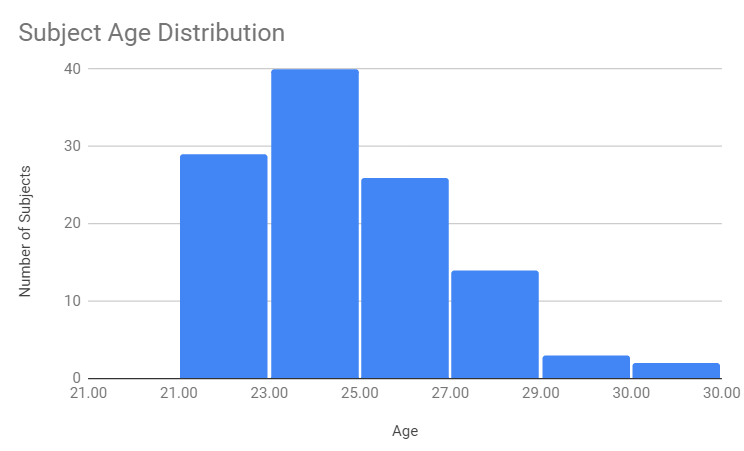}
\center \caption{Age histogram of 114 volunteers}
\label{FigureIBSDP_age}
\end{figure}

MR scans were collected from 114 young (21-30 years) healthy adult volunteers. All the adults had completed their schooling and a majority of (> 90$\%$) them also had an undergraduate level education. Healthy volunteers who had no past history of head injury were selected for the study. Figure\ref{FigureIBSDP_age} shows the age distribution of all the volunteers. An experienced psychiatrist examined all the volunteers and helped select only psychologically healthy subjects for the study. A clinician and an experienced neurologist examined all the scans to identify and exclude those with any structural abnormalities. Scans of 58 male and 56 female volunteer were selected finally for inclusion in our study after the scrutiny. Written consent was taken from all the volunteers for use of their anonymized MRI scan for research purposes. 
Scanning was done at three different sites which had different models of scanners as follows. The sitewise distribution was as follows: 39 subjects were scanned using Siemens 1.5T MRI scanner with T1 MPRAGE sequence, TE/TR/TI =  2.9 / 2370/ 1000  ms and flip angle=7$^{\circ}$; 38 subjects were scanned using GE 1.5T MRI scanner with T1 BRAVO sequence TE/TR/TI =  4.2 /10.2/450 ms and flip angle=15$^{\circ}$; and finally 37 subjects were scanned using Phillips 1.5T MRI scanner with T1 3D TFE sequence with TE/TR/TI = 3.8/8.2/- and flip angle=7$^{\circ}$. The imaging protocol was fixed to obtain scans with a voxel size of $1\times 1\times 1mm^3$ and a 3D matrix size of $256\times 256\times 192$. The MRI volumes were acquired using 192 sagittal cuts.

\subsection*{Preprocessing}
All the MRI volumes were pre-processed using a standard pipeline consisting of N4 Bias field correction \cite{N4bias} followed by denoising using Non-local Means filtering \cite{NLM}. Skull stripping was done with the Brain Extraction Tool\cite{BET} and all the images were checked manually slice by slice (using ITK-SNAP) to ensure good image quality after preprocessing.

\subsection*{Sub-cortical segmentation labels}
\begin{figure}
\centering
\includegraphics[scale=0.7]{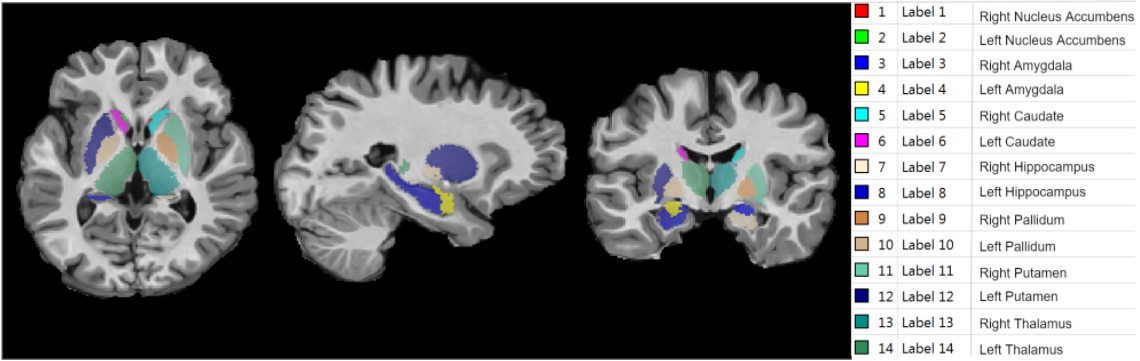}
\center \caption{Sub-cortical Structure labels of a sample subject image visualized in 3 planes. The labels corresponding to each color mapped structure is given on the right side. }
\label{FigureIBSDP_label}
\end{figure}

The seven sub-cortical structure pairs (Left and Right) chosen for markings were the Thalamus, Putamen, Pallidum, Hippocampus, Amygdala, Caudate and Accumbens area. All the manual markings were done by experts from Sree Chitra Tirunal Institute for Medical Sciences and Technology, India. The seven structures in each of the 2 hemispheres are illustrated with different colours in 3 canonical views for a sample slice in Figure \ref{FigureIBSDP_label}. 
\subsection*{Sub-cortical structure segmentation approach}
The process of image marking/labeling had 4 sequential steps: 
i) automated labeling, ii) label correction by a trained person, iii) label correction by a radiologist and iv) label finalization by a senior neuroradiologist. In order to perform the automated labeling, a set of 14 sub-cortical labels from Talairach Daemon labels \cite{Talairach} were transferred to each brain MRI scan. These were then manually edited using the ITK-SNAP \cite{itk} tool by a trained person to correct for overshoots and filling the missing voxels. The first two steps helped the radiologists to concentrate on fine tuning the markings and producing good quality labels. A radiologist then corrected the labels by overlaying them on the actual 3D MRI scans slice by slice. The corrected labels were checked in the 3 canonical views (coronal, sagittal and axial views) to verify the completeness of the 3D shape and labels. In the same way the corrected labels were finalized by a senior neruoradiologist. Experts delineated the structures using tissue intensity, relative position, and structure shape information from their experience. 3D-mesh visualization of each structure helped the experts to verify the delineated structure shape with the expected shape in their mind. The slice-by-slice delineation and 3D visualized cross-verification in each step helped the experts to work efficiently.

\section*{Data Records}
All the 114 scans and their labels are made publicly available at \url{https://doi.org/10.5281/zenodo.5656776}. The IBSD dataset is organised as follows: T1 weighted preprocessed 3D MRI scans and corresponding label files are stored in the main directory. Each label file has \textit{.ch.nii.gz} extension with same filename as the corresponding 3D MRI scan. All the image files are stored in NIFTI format. The sub-cortical structure labels are numbered  as given in Figure \ref{FigureIBSDP_label}.

\section*{Technical Validation}

The data quality of IBSD was checked by doing a comparison with other datasets. This is first done via a visual comparison of the image and the segmented structures. Next, IBSD data was used to train and test currently popular non-DL algorithms as well as state of the art DL algorithms which report results on public datasets. Specifically, segmentation was done with two popular toolboxes namely  the Freesurfer \cite{freesurfer} and FIRST \cite{first}; It was also done with DL methods namely 3D U-net \cite{3Dunet}, Residual 3D U-net  \cite{resunet}, Dense U-net \cite{dense} V-net \cite{vnet}, M-net \cite{rag} and the state of the art $\psi$-net \cite{psi}. The segmentation performance of Freesurfer \cite{freesurfer}, FIRST \cite{first}; 3D U-net \cite{3Dunet}, and $\psi$-net \cite{psi}
was assessed on IBSD as well as two other public datasets namely MICCAI 2012 \cite{MICCAIdataset} and IBSR \cite{ibsr} for comparison.

\subsubsection*{Evaluation Measures}
Two commonly used metrics are used for the quantitative evaluation of the segmentation methods. These are the Dice Similarity Coefficient (Dice) \cite{dice1945measures} and the Hausdorff Distance (HD) \cite{HD}. Dice helps assess the degree of overlap between the ground truth and computed segments while the HD helps  capture any tendency to over/under segment at a local level by a method. Since these two metrics help assess the accuracy of a method at a global and local level they are appropriate to evaluate  accurate and spatially consistency of segmentation results.
Let A and B be the predicted segmentation and ground truth respectively.
The Dice coefficient is found by computing the overlap between the computed segmentation result and the ground truth:
\begin{equation}
Dice(A,B) = \frac{2 \times |A \cap B|}{|A| + |B|},
\end{equation} 
Where $|A \cap B|$ denotes the number of pixels in the overlapping region between computed segmentation and ground truth while $|A| + |B|$ denotes the number of pixels in A and B. The Dice score varies between 0 and 1, with 0 indicating no overlap or  segmentation failure and 1 indicating complete overlap with ground truth or perfect segmentation.

HD is a spatial distance based metric, unlike Dice which assesses the overlap. HD therefore is based on computing the Euclidean distance between A and B as well as B and A as follows.
\begin{equation}
    HD(A,B) = max( h(A,B) , h(B,A) ),
\end{equation}
\begin{equation}
    h(A,B) = \max_{a\in A}\min_{b\in B} \parallel a - b \parallel,
\end{equation}
$\parallel.\parallel$ is the Euclidean distance. Unlike Dice, $HD$ is not bounded, however, lower values of $HD$ indicate better segmentation.

\begin{table}[]
\centering
\begin{tabular}{|c|c|c|c|c|c|}
\hline
Dataset    & Image Resolution                                                     & Age Range    & \# of subjects (Male:Female) & \begin{tabular}[c]{@{}c@{}}
\# of labels \\(total: sub-cortical) \\ \end{tabular} 
\\ \hline
IBSR       & \begin{tabular}[c]{@{}c@{}}0.93x0.93x1.5\\ or 1x1x1.5 $mm^3$\end{tabular} & Juvenile to 71 & 14:4        & 43:14                                                                                                                            \\ \hline
MICCAI2012 & 1x1x1.25 $mm^3$                                                         & 18-90        & 22:13       & 134:14                                                                                                                                                                        \\ \hline
LPBA40     & 2x2x2 $mm^3$                                                            & 19.3-39.5    & 20:20       & 56:6                                                                                                                                                                           \\ \hline
IBSD        & 1x1x1$mm^3$                                                             & 21-30        & 58:56       & 14:14                                                                                                                                                                           \\ \hline
\end{tabular}
\caption{Broad description of neuroimage datasets (of 1.5 T scans) for segmentation.}  
\label{Table:features}
\end{table}
\subsection*{Visual assessment of IBSD data}
\begin{figure}
\centering
\includegraphics[scale=0.7]{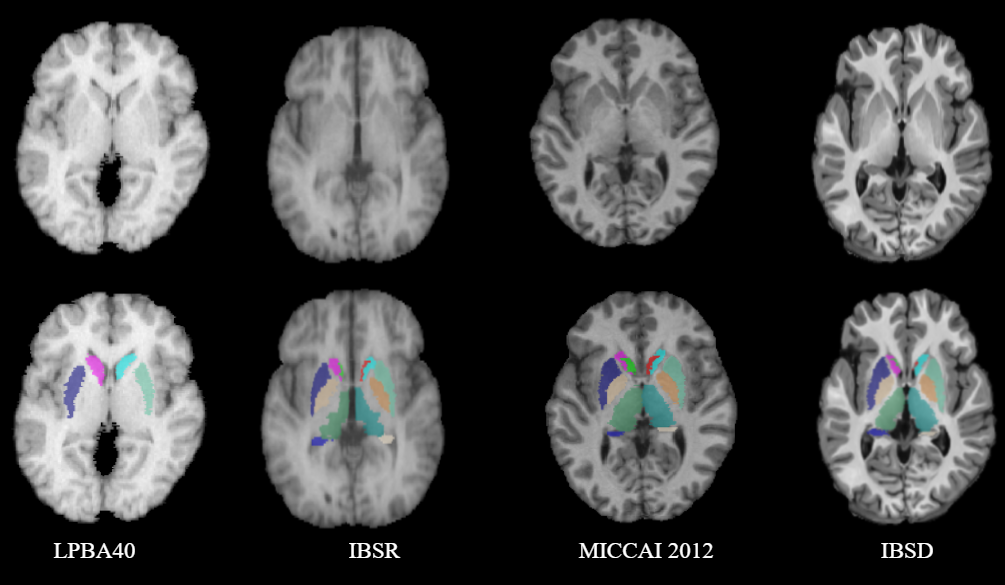}
\center \caption{Visual Comparison of  quality of MR image in (first row) and segmentation (second row) with central slices of the volume} 
\label{quality}
\end{figure}

Image resolution and quality have a major role in determining the quality of final manual delineations. The specifications of various datasets such as image quality parameters and subject information are listed for comparison in Table \ref{Table:features}. It can observed that IBSD has the highest number of scans (58+56 = 114) and the best image resolution (1mm isotropic voxel). A sample slice image from 3 public datasets (LPBA40, IBSR, MICCAI 2012) and IBSD is shown in Fig. \ref{quality}. The IBSD image is seen to be visually similar in the MICCAI 2012 dataset due to the similarity in image resolution, though with better contrast.
\subsection*{Validation of IBSD via automatic segmentation}
Validation of IBSD was done using two types of automated segmentation algorithms, namely those based on traditional machine learning and DL. 
If DL methods demonstrate good segmentation performance then it can be inferred that the size of the dataset and image quality are good as these DL methods are data driven.\\
The performance figures of two commonly used conventional segmentation tools, namely the Freesurfer \cite{freesurfer} and FIRST \cite{first}, are presented in TABLE \ref{table:conv} while those for DL-based methods are presented in Table \ref{table:segperf}. The Nucleus Accumbens (Label 1-2) is significantly smaller than others which affects the automatic segmentation performance. So, the performance  excluding Accumbens', i.e. only for labels 3-14, is also reported in TABLE \ref{table:conv} and \ref{table:segperf}.  \\
A six-fold cross validation was done to assess the DL methods by splitting the 114 volumes into six folds with 19 images in each; of these, four folds were used for training, one fold for validation and one for testing. Six different models were thus obtained, tested separately and the performance scores were averaged and reported for each segmentation method.\\ When one compares the results in Table \ref{table:segperf} with those for non-DL methods in TABLE \ref{table:conv}, it is apparent that the DL methods outperform the non-DL methods. Among the DL methods, the Dense U-net shows the best performance on IBSD, with a $14\%$ improvement in Dice and  nearly 100\% improvement in HD over the best non-DL (FIRST) method.
Exclusion/inclusion of Accumbens appears to impact the Dice score but not the HD value as indicated by the figures in the last 2 rows of the Table. A consistent performance is observed with DL methods (see Table \ref{table:segperf}) using U-net based architectures on the IBSD dataset as the Dice scores are between 0.87 to 0.88 and HD value is between 2.9 and 5. This attests to the integrity of the IBSD data. 
\begin{table}[h]
\begin{tabular}{|c|c|c|c|l|l|}
\hline
Label & Structure               & \multicolumn{2}{c|}{Freesurfer}                                                                                                              & \multicolumn{2}{c|}{FIRST}                                                                                                                                                 \\ \hline
      &                         & \cellcolor[HTML]{FFCCC9}Dice                         & \cellcolor[HTML]{FFFFC7}\begin{tabular}[c]{@{}c@{}}HD\end{tabular} & \multicolumn{1}{c|}{\cellcolor[HTML]{FFCCC9}Dice}             & \multicolumn{1}{c|}{\cellcolor[HTML]{FFFFC7}\begin{tabular}[c]{@{}c@{}}HD\end{tabular}} \\ \hline
1     & Right Nucleus Accumbens & \cellcolor[HTML]{FFCCC9}0.55 $\pm$ 0.08 & \cellcolor[HTML]{FFFFC7}7.14 $\pm$ 3.71                                  & \cellcolor[HTML]{FFCCC9}0.64 $\pm$ 0.09          & \cellcolor[HTML]{FFFFC7}6.27 $\pm$ 3.45                                                       \\ \hline
2     & Left Nucleus Accumbens  & \cellcolor[HTML]{FFCCC9}0.51 $\pm$ 0.13 & \cellcolor[HTML]{FFFFC7}8.67 $\pm$ 3.37                                  & \cellcolor[HTML]{FFCCC9}0.57 $\pm$ 0.09          & \cellcolor[HTML]{FFFFC7}16.26  $\pm$ 10.42                                                    \\ \hline
3     & Right Amygdala          & \cellcolor[HTML]{FFCCC9}0.66 $\pm$ 0.04 & \cellcolor[HTML]{FFFFC7}5.68 $\pm$ 4.7                                   & \cellcolor[HTML]{FFCCC9}0.71 $\pm$  0.05         & \cellcolor[HTML]{FFFFC7}4.9 $\pm$ 4.82                                                        \\ \hline
4     & Left Amygdala           & \cellcolor[HTML]{FFCCC9}0.67 $\pm$ 0.05 & \cellcolor[HTML]{FFFFC7}4.16 $\pm$ 0.8                                   & \cellcolor[HTML]{FFCCC9}0.71 $\pm$  0.05         & \cellcolor[HTML]{FFFFC7}3.76  $\pm$ 1.06                                                      \\ \hline
5     & Right Caudate           & \cellcolor[HTML]{FFCCC9}0.8 $\pm$ 0.03  & \cellcolor[HTML]{FFFFC7}8.02  $\pm$ 2.64                                 & \cellcolor[HTML]{FFCCC9}0.79 $\pm$  0.04         & \cellcolor[HTML]{FFFFC7}5.31  $\pm$  1.38                                                     \\ \hline
6     & Left Caudate            & \cellcolor[HTML]{FFCCC9}0.81 $\pm$ 0.03 & \cellcolor[HTML]{FFFFC7}7.83  $\pm$ 2.48                                 & \cellcolor[HTML]{FFCCC9}0.79 $\pm$  0.05         & \cellcolor[HTML]{FFFFC7}4.36  $\pm$  1.43                                                     \\ \hline
7     & Right Hippocampus       & \cellcolor[HTML]{FFCCC9}0.81 $\pm$ 0.03 & \cellcolor[HTML]{FFFFC7}4.18  $\pm$  0.81                                & \cellcolor[HTML]{FFCCC9}0.8 $\pm$  0.03          & \cellcolor[HTML]{FFFFC7}3.96  $\pm$ 0.89                                                      \\ \hline
8     & Left Hippocampus        & \cellcolor[HTML]{FFCCC9}0.77 $\pm$ 0.03 & \cellcolor[HTML]{FFFFC7}5.04  $\pm$ 0.96                                 & \cellcolor[HTML]{FFCCC9}0.76 $\pm$ 0.04          & \cellcolor[HTML]{FFFFC7}4.78  $\pm$  1.28                                                     \\ \hline
9     & Right Pallidum          & \cellcolor[HTML]{FFCCC9}0.77 $\pm$ 0.03 & \cellcolor[HTML]{FFFFC7}4.44  $\pm$  1.2                                 & \cellcolor[HTML]{FFCCC9}0.79 $\pm$ 0.05          & \cellcolor[HTML]{FFFFC7}3.43  $\pm$  0.67                                                     \\ \hline
10    & Left Pallidum           & \cellcolor[HTML]{FFCCC9}0.69 $\pm$ 0.03 & \cellcolor[HTML]{FFFFC7}4.39  $\pm$  1.19                                & \cellcolor[HTML]{FFCCC9}0.78 $\pm$ 0.06          & \cellcolor[HTML]{FFFFC7}3.04  $\pm$  0.61                                                     \\ \hline
11    & Right Putamen           & \cellcolor[HTML]{FFCCC9}0.83 $\pm$ 0.03 & \cellcolor[HTML]{FFFFC7}5.31  $\pm$ 1.84                                 & \cellcolor[HTML]{FFCCC9}0.85 $\pm$ 0.03          & \cellcolor[HTML]{FFFFC7}4.95  $\pm$ 1.98                                                      \\ \hline
12    & Left Putamen            & \cellcolor[HTML]{FFCCC9}0.82 $\pm$ 0.05 & \cellcolor[HTML]{FFFFC7}8.54  $\pm$  2.84                                & \cellcolor[HTML]{FFCCC9}0.85 $\pm$ 0.03          & \cellcolor[HTML]{FFFFC7}8.56 $\pm$  2.88                                                      \\ \hline
13    & Right Thalamus          & \cellcolor[HTML]{FFCCC9}0.84 $\pm$ 0.02 & \cellcolor[HTML]{FFFFC7}4.69  $\pm$  0.53                                & \cellcolor[HTML]{FFCCC9}0.86 $\pm$ 0.02          & \cellcolor[HTML]{FFFFC7}4.47  $\pm$  0.7                                                      \\ \hline
14    & Left Thalamus           & \cellcolor[HTML]{FFCCC9}0.83 $\pm$ 0.03 & \cellcolor[HTML]{FFFFC7}4.75  $\pm$  0.89                                & \cellcolor[HTML]{FFCCC9}0.86 $\pm$ 0.02          & \cellcolor[HTML]{FFFFC7}4.22  $\pm$  0.81                                                     \\ \hline
      & Average of label 3-14   & \cellcolor[HTML]{FFCCC9}0.78 $\pm$ 0.04 & \cellcolor[HTML]{FFFFC7}5.59  $\pm$  1.74                                & \cellcolor[HTML]{FFCCC9}\textbf{0.8 $\pm$ 0.04}  & \cellcolor[HTML]{FFFFC7}\textbf{4.65  $\pm$  1.54}                                            \\ \hline
      & Full Average            & \cellcolor[HTML]{FFCCC9}0.74 $\pm$ 0.05 & \cellcolor[HTML]{FFFFC7}5.92  $\pm$  2                                   & \cellcolor[HTML]{FFCCC9}\textbf{0.77 $\pm$ 0.05} & \cellcolor[HTML]{FFFFC7}\textbf{5.59 $\pm$ 2.31}                                              \\ \hline
\end{tabular}
\caption{Segmenttation performance of Freesurfer and FIRST on IBSD data. The mean Mean$\pm$ Standard Deviation values are reported for Dice coefficient and HD.}
\label{table:conv}
\end{table}

\begin{table}[ht]
\hfil
\begin{sideways}
\begin{minipage}{1\textheight}
\begin{tabular}{|c|c|l|l|l|l|l|l|l|l|l|l|l|l|l|l|l|l|l|l|}
\hline
Label                  & Structure                                                            & \multicolumn{3}{c|}{3D U-net}                                                                                                                                                                   & \multicolumn{3}{c|}{\begin{tabular}[c]{@{}c@{}}Residual 3D \\ U-net\end{tabular}}                                                                                                               & \multicolumn{3}{c|}{Dense U-net}                                                                                                                                                                      & \multicolumn{3}{c|}{V-net}                                                                                                                                                                      & \multicolumn{3}{l|}{M-net}                                                                                                                                                                      & \multicolumn{3}{c|}{$\psi-net$}                                                                                                                                                                          \\ \hline
                       &                                                                      & \multicolumn{1}{c|}{\cellcolor[HTML]{FFCCC9}Dice}                                  & \multicolumn{2}{c|}{\cellcolor[HTML]{FFFFC7}\begin{tabular}[c]{@{}c@{}}HD\end{tabular}} & \multicolumn{1}{c|}{\cellcolor[HTML]{FFCCC9}Dice}                                  & \multicolumn{2}{c|}{\cellcolor[HTML]{FFFFC7}\begin{tabular}[c]{@{}c@{}}HD\end{tabular}} & \multicolumn{1}{c|}{\cellcolor[HTML]{FFCCC9}Dice}                                  & \multicolumn{2}{c|}{\cellcolor[HTML]{FFFFC7}\begin{tabular}[c]{@{}c@{}}HD\end{tabular}}       & \multicolumn{1}{c|}{\cellcolor[HTML]{FFCCC9}Dice}                                  & \multicolumn{2}{c|}{\cellcolor[HTML]{FFFFC7}\begin{tabular}[c]{@{}c@{}}HD\end{tabular}} & \multicolumn{1}{c|}{\cellcolor[HTML]{FFCCC9}Dice}                                  & \multicolumn{2}{c|}{\cellcolor[HTML]{FFFFC7}\begin{tabular}[c]{@{}c@{}}HD\end{tabular}} & \multicolumn{1}{c|}{\cellcolor[HTML]{FFCCC9}Dice}                                           & \multicolumn{2}{c|}{\cellcolor[HTML]{FFFFC7}\begin{tabular}[c]{@{}c@{}}HD\end{tabular}} \\ \hline
1                      & \begin{tabular}[c]{@{}c@{}}Right\\ Nucleus \\ Accumbens\end{tabular} & \cellcolor[HTML]{FFCCC9}\begin{tabular}[c]{@{}l@{}}0.751\\ $\pm$0.06\end{tabular}  & \multicolumn{2}{l|}{\cellcolor[HTML]{FFFFC7}\begin{tabular}[c]{@{}l@{}}3.856\\ $\pm$0.147\end{tabular}}    & \cellcolor[HTML]{FFCCC9}\begin{tabular}[c]{@{}l@{}}0.767\\ $\pm$0.061\end{tabular} & \multicolumn{2}{l|}{\cellcolor[HTML]{FFFFC7}\begin{tabular}[c]{@{}l@{}}4.68\\ $\pm$1.827\end{tabular}}     & \cellcolor[HTML]{FFCCC9}\begin{tabular}[c]{@{}l@{}}0.778\\ $\pm$0.056\end{tabular} & \multicolumn{2}{l|}{\cellcolor[HTML]{FFFFC7}\begin{tabular}[c]{@{}l@{}}3.783\\ $\pm$0.412\end{tabular}}          & \cellcolor[HTML]{FFCCC9}\begin{tabular}[c]{@{}l@{}}0.74\\ $\pm$0.068\end{tabular}  & \multicolumn{2}{l|}{\cellcolor[HTML]{FFFFC7}\begin{tabular}[c]{@{}l@{}}3.669\\ $\pm$0.628\end{tabular}}    & \cellcolor[HTML]{FFCCC9}\begin{tabular}[c]{@{}l@{}}0.788\\ $\pm$0.012\end{tabular} & \multicolumn{2}{l|}{\cellcolor[HTML]{FFFFC7}\begin{tabular}[c]{@{}l@{}}3.759\\ $\pm$0.472\end{tabular}}    & \cellcolor[HTML]{FFCCC9}\begin{tabular}[c]{@{}l@{}}0.782\\ $\pm$0.06\end{tabular}           & \multicolumn{2}{l|}{\cellcolor[HTML]{FFFFC7}\begin{tabular}[c]{@{}l@{}}3.893\\ $\pm$0.373\end{tabular}}    \\ \hline
2                      & \begin{tabular}[c]{@{}c@{}}Left \\ Nucleus \\ Accumbens\end{tabular} & \cellcolor[HTML]{FFCCC9}\begin{tabular}[c]{@{}l@{}}0.708\\ $\pm$0.056\end{tabular} & \multicolumn{2}{l|}{\cellcolor[HTML]{FFFFC7}\begin{tabular}[c]{@{}l@{}}4.026\\ $\pm$0.377\end{tabular}}    & \cellcolor[HTML]{FFCCC9}\begin{tabular}[c]{@{}l@{}}0.72\\ $\pm$0.052\end{tabular}  & \multicolumn{2}{l|}{\cellcolor[HTML]{FFFFC7}\begin{tabular}[c]{@{}l@{}}3.706\\ $\pm$0.244\end{tabular}}    & \cellcolor[HTML]{FFCCC9}\begin{tabular}[c]{@{}l@{}}0.728\\ $\pm$0.054\end{tabular} & \multicolumn{2}{l|}{\cellcolor[HTML]{FFFFC7}\begin{tabular}[c]{@{}l@{}}3.868\\ $\pm$0.15\end{tabular}}           & \cellcolor[HTML]{FFCCC9}\begin{tabular}[c]{@{}l@{}}0.7\\ $\pm$0.057\end{tabular}   & \multicolumn{2}{l|}{\cellcolor[HTML]{FFFFC7}\begin{tabular}[c]{@{}l@{}}3.865\\ $\pm$0.475\end{tabular}}    & \cellcolor[HTML]{FFCCC9}\begin{tabular}[c]{@{}l@{}}0.735\\ $\pm$0.008\end{tabular} & \multicolumn{2}{l|}{\cellcolor[HTML]{FFFFC7}\begin{tabular}[c]{@{}l@{}}3.769\\ $\pm$0.215\end{tabular}}    & \cellcolor[HTML]{FFCCC9}\begin{tabular}[c]{@{}l@{}}0.729\\ $\pm$0.061\end{tabular}          & \multicolumn{2}{l|}{\cellcolor[HTML]{FFFFC7}\begin{tabular}[c]{@{}l@{}}3.955\\ $\pm$0.308\end{tabular}}    \\ \hline
3                      & \begin{tabular}[c]{@{}c@{}}Right\\  Amygdala\end{tabular}            & \cellcolor[HTML]{FFCCC9}\begin{tabular}[c]{@{}l@{}}0.849\\ $\pm$0.028\end{tabular} & \multicolumn{2}{l|}{\cellcolor[HTML]{FFFFC7}\begin{tabular}[c]{@{}l@{}}3.038\\ $\pm$0.858\end{tabular}}    & \cellcolor[HTML]{FFCCC9}\begin{tabular}[c]{@{}l@{}}0.86\\ $\pm$0.031\end{tabular}  & \multicolumn{2}{l|}{\cellcolor[HTML]{FFFFC7}\begin{tabular}[c]{@{}l@{}}3.499\\ $\pm$1.224\end{tabular}}    & \cellcolor[HTML]{FFCCC9}\begin{tabular}[c]{@{}l@{}}0.864\\ $\pm$0.029\end{tabular} & \multicolumn{2}{l|}{\cellcolor[HTML]{FFFFC7}\begin{tabular}[c]{@{}l@{}}2.915\\ $\pm$0.842\end{tabular}}          & \cellcolor[HTML]{FFCCC9}\begin{tabular}[c]{@{}l@{}}0.839\\ $\pm$0.038\end{tabular} & \multicolumn{2}{l|}{\cellcolor[HTML]{FFFFC7}\begin{tabular}[c]{@{}l@{}}3.378\\ $\pm$1.112\end{tabular}}    & \cellcolor[HTML]{FFCCC9}\begin{tabular}[c]{@{}l@{}}0.872\\ $\pm$0.004\end{tabular} & \multicolumn{2}{l|}{\cellcolor[HTML]{FFFFC7}\begin{tabular}[c]{@{}l@{}}2.884\\ $\pm$0.733\end{tabular}}    & \cellcolor[HTML]{FFCCC9}\begin{tabular}[c]{@{}l@{}}0.867\\ $\pm$0.036\end{tabular}          & \multicolumn{2}{l|}{\cellcolor[HTML]{FFFFC7}\begin{tabular}[c]{@{}l@{}}3.475\\ $\pm$1.472\end{tabular}}    \\ \hline
4                      & \begin{tabular}[c]{@{}c@{}}Left \\ Amygdala\end{tabular}             & \cellcolor[HTML]{FFCCC9}\begin{tabular}[c]{@{}l@{}}0.828\\ $\pm$0.03\end{tabular}  & \multicolumn{2}{l|}{\cellcolor[HTML]{FFFFC7}\begin{tabular}[c]{@{}l@{}}3.195\\ $\pm$0.299\end{tabular}}    & \cellcolor[HTML]{FFCCC9}\begin{tabular}[c]{@{}l@{}}0.841\\ $\pm$0.035\end{tabular} & \multicolumn{2}{l|}{\cellcolor[HTML]{FFFFC7}\begin{tabular}[c]{@{}l@{}}3.423\\ $\pm$0.85\end{tabular}}     & \cellcolor[HTML]{FFCCC9}\begin{tabular}[c]{@{}l@{}}0.846\\ $\pm$0.031\end{tabular} & \multicolumn{2}{l|}{\cellcolor[HTML]{FFFFC7}\begin{tabular}[c]{@{}l@{}}2.879\\ $\pm$0.358\end{tabular}}          & \cellcolor[HTML]{FFCCC9}\begin{tabular}[c]{@{}l@{}}0.805\\ $\pm$0.054\end{tabular} & \multicolumn{2}{l|}{\cellcolor[HTML]{FFFFC7}\begin{tabular}[c]{@{}l@{}}3.252\\ $\pm$0.54\end{tabular}}     & \cellcolor[HTML]{FFCCC9}\begin{tabular}[c]{@{}l@{}}0.854\\ $\pm$0.004\end{tabular} & \multicolumn{2}{l|}{\cellcolor[HTML]{FFFFC7}\begin{tabular}[c]{@{}l@{}}2.958\\ $\pm$0.63\end{tabular}}     & \cellcolor[HTML]{FFCCC9}\begin{tabular}[c]{@{}l@{}}0.849\\ $\pm$0.039\end{tabular}          & \multicolumn{2}{l|}{\cellcolor[HTML]{FFFFC7}\begin{tabular}[c]{@{}l@{}}3.074\\ $\pm$0.379\end{tabular}}    \\ \hline
5                      & \begin{tabular}[c]{@{}c@{}}Right \\ Caudate\end{tabular}             & \cellcolor[HTML]{FFCCC9}\begin{tabular}[c]{@{}l@{}}0.902\\ $\pm$0.015\end{tabular} & \multicolumn{2}{l|}{\cellcolor[HTML]{FFFFC7}\begin{tabular}[c]{@{}l@{}}3.11\\ $\pm$0.849\end{tabular}}     & \cellcolor[HTML]{FFCCC9}\begin{tabular}[c]{@{}l@{}}0.905\\ $\pm$0.017\end{tabular} & \multicolumn{2}{l|}{\cellcolor[HTML]{FFFFC7}\begin{tabular}[c]{@{}l@{}}3.512\\ $\pm$0.926\end{tabular}}    & \cellcolor[HTML]{FFCCC9}\begin{tabular}[c]{@{}l@{}}0.908\\ $\pm$0.014\end{tabular} & \multicolumn{2}{l|}{\cellcolor[HTML]{FFFFC7}\begin{tabular}[c]{@{}l@{}}2.911\\ $\pm$0.707\end{tabular}}          & \cellcolor[HTML]{FFCCC9}\begin{tabular}[c]{@{}l@{}}0.896\\ $\pm$0.031\end{tabular} & \multicolumn{2}{l|}{\cellcolor[HTML]{FFFFC7}\begin{tabular}[c]{@{}l@{}}2.718\\ $\pm$0.311\end{tabular}}    & \cellcolor[HTML]{FFCCC9}\begin{tabular}[c]{@{}l@{}}0.903\\ $\pm$0.003\end{tabular} & \multicolumn{2}{l|}{\cellcolor[HTML]{FFFFC7}\begin{tabular}[c]{@{}l@{}}2.864\\ $\pm$0.984\end{tabular}}    & \cellcolor[HTML]{FFCCC9}\begin{tabular}[c]{@{}l@{}}0.909\\ $\pm$0.015\end{tabular}          & \multicolumn{2}{l|}{\cellcolor[HTML]{FFFFC7}\begin{tabular}[c]{@{}l@{}}7.952\\ $\pm$8.141\end{tabular}}    \\ \hline
6                      & \begin{tabular}[c]{@{}c@{}}Left \\ Caudate\end{tabular}              & \cellcolor[HTML]{FFCCC9}\begin{tabular}[c]{@{}l@{}}0.902\\ $\pm$0.014\end{tabular} & \multicolumn{2}{l|}{\cellcolor[HTML]{FFFFC7}\begin{tabular}[c]{@{}l@{}}3.83\\ $\pm$1.854\end{tabular}}     & \cellcolor[HTML]{FFCCC9}\begin{tabular}[c]{@{}l@{}}0.906\\ $\pm$0.014\end{tabular} & \multicolumn{2}{l|}{\cellcolor[HTML]{FFFFC7}\begin{tabular}[c]{@{}l@{}}3.91\\ $\pm$0.487\end{tabular}}     & \cellcolor[HTML]{FFCCC9}\begin{tabular}[c]{@{}l@{}}0.908\\ $\pm$0.013\end{tabular} & \multicolumn{2}{l|}{\cellcolor[HTML]{FFFFC7}\begin{tabular}[c]{@{}l@{}}2.96\\ $\pm$0.158\end{tabular}}           & \cellcolor[HTML]{FFCCC9}\begin{tabular}[c]{@{}l@{}}0.898\\ $\pm$0.022\end{tabular} & \multicolumn{2}{l|}{\cellcolor[HTML]{FFFFC7}\begin{tabular}[c]{@{}l@{}}3.647\\ $\pm$1.319\end{tabular}}    & \cellcolor[HTML]{FFCCC9}\begin{tabular}[c]{@{}l@{}}0.903\\ $\pm$0.002\end{tabular} & \multicolumn{2}{l|}{\cellcolor[HTML]{FFFFC7}\begin{tabular}[c]{@{}l@{}}3.234\\ $\pm$1.073\end{tabular}}    & \cellcolor[HTML]{FFCCC9}\begin{tabular}[c]{@{}l@{}}0.909\\ $\pm$0.015\end{tabular}          & \multicolumn{2}{l|}{\cellcolor[HTML]{FFFFC7}\begin{tabular}[c]{@{}l@{}}4.396\\ $\pm$1.216\end{tabular}}    \\ \hline
7                      & \begin{tabular}[c]{@{}c@{}}Right \\ Hippocampus\end{tabular}         & \cellcolor[HTML]{FFCCC9}\begin{tabular}[c]{@{}l@{}}0.885\\ $\pm$0.013\end{tabular} & \multicolumn{2}{l|}{\cellcolor[HTML]{FFFFC7}\begin{tabular}[c]{@{}l@{}}3.884\\ $\pm$1.236\end{tabular}}    & \cellcolor[HTML]{FFCCC9}\begin{tabular}[c]{@{}l@{}}0.891\\ $\pm$0.012\end{tabular} & \multicolumn{2}{l|}{\cellcolor[HTML]{FFFFC7}\begin{tabular}[c]{@{}l@{}}3.903\\ $\pm$2.283\end{tabular}}    & \cellcolor[HTML]{FFCCC9}\begin{tabular}[c]{@{}l@{}}0.894\\ $\pm$0.011\end{tabular} & \multicolumn{2}{l|}{\cellcolor[HTML]{FFFFC7}\begin{tabular}[c]{@{}l@{}}2.688\\ $\pm$0.265\end{tabular}}          & \cellcolor[HTML]{FFCCC9}\begin{tabular}[c]{@{}l@{}}0.875\\ $\pm$0.02\end{tabular}  & \multicolumn{2}{l|}{\cellcolor[HTML]{FFFFC7}\begin{tabular}[c]{@{}l@{}}3.584\\ $\pm$0.717\end{tabular}}    & \cellcolor[HTML]{FFCCC9}\begin{tabular}[c]{@{}l@{}}0.889\\ $\pm$0.004\end{tabular} & \multicolumn{2}{l|}{\cellcolor[HTML]{FFFFC7}\begin{tabular}[c]{@{}l@{}}3.361\\ $\pm$1.103\end{tabular}}    & \cellcolor[HTML]{FFCCC9}\begin{tabular}[c]{@{}l@{}}0.898\\ $\pm$0.012\end{tabular}          & \multicolumn{2}{l|}{\cellcolor[HTML]{FFFFC7}\begin{tabular}[c]{@{}l@{}}7.533\\ $\pm$3.66\end{tabular}}     \\ \hline
8                      & \begin{tabular}[c]{@{}c@{}}Left\\  Hippocampus\end{tabular}          & \cellcolor[HTML]{FFCCC9}\begin{tabular}[c]{@{}l@{}}0.871\\ $\pm$0.017\end{tabular} & \multicolumn{2}{l|}{\cellcolor[HTML]{FFFFC7}\begin{tabular}[c]{@{}l@{}}3.715\\ $\pm$1.516\end{tabular}}    & \cellcolor[HTML]{FFCCC9}\begin{tabular}[c]{@{}l@{}}0.876\\ $\pm$0.017\end{tabular} & \multicolumn{2}{l|}{\cellcolor[HTML]{FFFFC7}\begin{tabular}[c]{@{}l@{}}3.382\\ $\pm$0.548\end{tabular}}    & \cellcolor[HTML]{FFCCC9}\begin{tabular}[c]{@{}l@{}}0.88\\ $\pm$0.016\end{tabular}  & \multicolumn{2}{l|}{\cellcolor[HTML]{FFFFC7}\begin{tabular}[c]{@{}l@{}}3.086\\ $\pm$0.479\end{tabular}}          & \cellcolor[HTML]{FFCCC9}\begin{tabular}[c]{@{}l@{}}0.858\\ $\pm$0.029\end{tabular} & \multicolumn{2}{l|}{\cellcolor[HTML]{FFFFC7}\begin{tabular}[c]{@{}l@{}}3.28\\ $\pm$0.305\end{tabular}}     & \cellcolor[HTML]{FFCCC9}\begin{tabular}[c]{@{}l@{}}0.874\\ $\pm$0.003\end{tabular} & \multicolumn{2}{l|}{\cellcolor[HTML]{FFFFC7}\begin{tabular}[c]{@{}l@{}}3.709\\ $\pm$1.314\end{tabular}}    & \cellcolor[HTML]{FFCCC9}\begin{tabular}[c]{@{}l@{}}0.883\\ $\pm$0.019\end{tabular}          & \multicolumn{2}{l|}{\cellcolor[HTML]{FFFFC7}\begin{tabular}[c]{@{}l@{}}6.934\\ $\pm$3.323\end{tabular}}    \\ \hline
9                      & \begin{tabular}[c]{@{}c@{}}Right \\ Pallidum\end{tabular}            & \cellcolor[HTML]{FFCCC9}\begin{tabular}[c]{@{}l@{}}0.88\\ $\pm$0.017\end{tabular}  & \multicolumn{2}{l|}{\cellcolor[HTML]{FFFFC7}\begin{tabular}[c]{@{}l@{}}2.021\\ $\pm$0.101\end{tabular}}    & \cellcolor[HTML]{FFCCC9}\begin{tabular}[c]{@{}l@{}}0.887\\ $\pm$0.017\end{tabular} & \multicolumn{2}{l|}{\cellcolor[HTML]{FFFFC7}\begin{tabular}[c]{@{}l@{}}2.983\\ $\pm$1.996\end{tabular}}    & \cellcolor[HTML]{FFCCC9}\begin{tabular}[c]{@{}l@{}}0.893\\ $\pm$0.015\end{tabular} & \multicolumn{2}{l|}{\cellcolor[HTML]{FFFFC7}\begin{tabular}[c]{@{}l@{}}1.811\\ $\pm$0.075\end{tabular}}          & \cellcolor[HTML]{FFCCC9}\begin{tabular}[c]{@{}l@{}}0.868\\ $\pm$0.021\end{tabular} & \multicolumn{2}{l|}{\cellcolor[HTML]{FFFFC7}\begin{tabular}[c]{@{}l@{}}2.275\\ $\pm$0.134\end{tabular}}    & \cellcolor[HTML]{FFCCC9}\begin{tabular}[c]{@{}l@{}}0.893\\ $\pm$0.003\end{tabular} & \multicolumn{2}{l|}{\cellcolor[HTML]{FFFFC7}\begin{tabular}[c]{@{}l@{}}2.106\\ $\pm$0.769\end{tabular}}    & \cellcolor[HTML]{FFCCC9}\begin{tabular}[c]{@{}l@{}}0.904\\ $\pm$0.014\end{tabular}          & \multicolumn{2}{l|}{\cellcolor[HTML]{FFFFC7}\begin{tabular}[c]{@{}l@{}}2.734\\ $\pm$2.176\end{tabular}}    \\ \hline
10                     & \begin{tabular}[c]{@{}c@{}}Left \\ Pallidum\end{tabular}             & \cellcolor[HTML]{FFCCC9}\begin{tabular}[c]{@{}l@{}}0.871\\ $\pm$0.022\end{tabular} & \multicolumn{2}{l|}{\cellcolor[HTML]{FFFFC7}\begin{tabular}[c]{@{}l@{}}2.121\\ $\pm$0.076\end{tabular}}    & \cellcolor[HTML]{FFCCC9}\begin{tabular}[c]{@{}l@{}}0.882\\ $\pm$0.021\end{tabular} & \multicolumn{2}{l|}{\cellcolor[HTML]{FFFFC7}\begin{tabular}[c]{@{}l@{}}2.029\\ $\pm$0.113\end{tabular}}    & \cellcolor[HTML]{FFCCC9}\begin{tabular}[c]{@{}l@{}}0.883\\ $\pm$0.021\end{tabular} & \multicolumn{2}{l|}{\cellcolor[HTML]{FFFFC7}\begin{tabular}[c]{@{}l@{}}2.01\\ $\pm$0.133\end{tabular}}           & \cellcolor[HTML]{FFCCC9}\begin{tabular}[c]{@{}l@{}}0.86\\ $\pm$0.027\end{tabular}  & \multicolumn{2}{l|}{\cellcolor[HTML]{FFFFC7}\begin{tabular}[c]{@{}l@{}}2.345\\ $\pm$0.189\end{tabular}}    & \cellcolor[HTML]{FFCCC9}\begin{tabular}[c]{@{}l@{}}0.886\\ $\pm$0.002\end{tabular} & \multicolumn{2}{l|}{\cellcolor[HTML]{FFFFC7}\begin{tabular}[c]{@{}l@{}}2.351\\ $\pm$0.707\end{tabular}}    & \cellcolor[HTML]{FFCCC9}\begin{tabular}[c]{@{}l@{}}0.895\\ $\pm$0.02\end{tabular}           & \multicolumn{2}{l|}{\cellcolor[HTML]{FFFFC7}\begin{tabular}[c]{@{}l@{}}2.176\\ $\pm$0.129\end{tabular}}    \\ \hline
11                     & \begin{tabular}[c]{@{}c@{}}Right \\ Putamen\end{tabular}             & \cellcolor[HTML]{FFCCC9}\begin{tabular}[c]{@{}l@{}}0.907\\ $\pm$0.013\end{tabular} & \multicolumn{2}{l|}{\cellcolor[HTML]{FFFFC7}\begin{tabular}[c]{@{}l@{}}3.213\\ $\pm$0.134\end{tabular}}    & \cellcolor[HTML]{FFCCC9}\begin{tabular}[c]{@{}l@{}}0.911\\ $\pm$0.013\end{tabular} & \multicolumn{2}{l|}{\cellcolor[HTML]{FFFFC7}\begin{tabular}[c]{@{}l@{}}3.783\\ $\pm$1.108\end{tabular}}    & \cellcolor[HTML]{FFCCC9}\begin{tabular}[c]{@{}l@{}}0.915\\ $\pm$0.014\end{tabular} & \multicolumn{2}{l|}{\cellcolor[HTML]{FFFFC7}\begin{tabular}[c]{@{}l@{}}3.206\\ $\pm$0.161\end{tabular}}          & \cellcolor[HTML]{FFCCC9}\begin{tabular}[c]{@{}l@{}}0.904\\ $\pm$0.02\end{tabular}  & \multicolumn{2}{l|}{\cellcolor[HTML]{FFFFC7}\begin{tabular}[c]{@{}l@{}}3.424\\ $\pm$0.509\end{tabular}}    & \cellcolor[HTML]{FFCCC9}\begin{tabular}[c]{@{}l@{}}0.908\\ $\pm$0.001\end{tabular} & \multicolumn{2}{l|}{\cellcolor[HTML]{FFFFC7}\begin{tabular}[c]{@{}l@{}}3.027\\ $\pm$0.564\end{tabular}}    & \cellcolor[HTML]{FFCCC9}\begin{tabular}[c]{@{}l@{}}0.92\\ $\pm$0.012\end{tabular}           & \multicolumn{2}{l|}{\cellcolor[HTML]{FFFFC7}\begin{tabular}[c]{@{}l@{}}4.511\\ $\pm$2.214\end{tabular}}    \\ \hline
12                     & \begin{tabular}[c]{@{}c@{}}Left \\ Putamen\end{tabular}              & \cellcolor[HTML]{FFCCC9}\begin{tabular}[c]{@{}l@{}}0.902\\ $\pm$0.014\end{tabular} & \multicolumn{2}{l|}{\cellcolor[HTML]{FFFFC7}\begin{tabular}[c]{@{}l@{}}3.747\\ $\pm$0.441\end{tabular}}    & \cellcolor[HTML]{FFCCC9}\begin{tabular}[c]{@{}l@{}}0.906\\ $\pm$0.014\end{tabular} & \multicolumn{2}{l|}{\cellcolor[HTML]{FFFFC7}\begin{tabular}[c]{@{}l@{}}3.671\\ $\pm$0.719\end{tabular}}    & \cellcolor[HTML]{FFCCC9}\begin{tabular}[c]{@{}l@{}}0.909\\ $\pm$0.013\end{tabular} & \multicolumn{2}{l|}{\cellcolor[HTML]{FFFFC7}\begin{tabular}[c]{@{}l@{}}4.126\\ $\pm$1.419\end{tabular}}          & \cellcolor[HTML]{FFCCC9}\begin{tabular}[c]{@{}l@{}}0.899\\ $\pm$0.02\end{tabular}  & \multicolumn{2}{l|}{\cellcolor[HTML]{FFFFC7}\begin{tabular}[c]{@{}l@{}}4.071\\ $\pm$1.358\end{tabular}}    & \cellcolor[HTML]{FFCCC9}\begin{tabular}[c]{@{}l@{}}0.903\\ $\pm$0.001\end{tabular} & \multicolumn{2}{l|}{\cellcolor[HTML]{FFFFC7}\begin{tabular}[c]{@{}l@{}}3.565\\ $\pm$0.496\end{tabular}}    & \cellcolor[HTML]{FFCCC9}\begin{tabular}[c]{@{}l@{}}0.913\\ $\pm$0.015\end{tabular}          & \multicolumn{2}{l|}{\cellcolor[HTML]{FFFFC7}\begin{tabular}[c]{@{}l@{}}6.973\\ $\pm$3.541\end{tabular}}    \\ \hline
13                     & \begin{tabular}[c]{@{}c@{}}Right \\ Thalamus\end{tabular}            & \cellcolor[HTML]{FFCCC9}\begin{tabular}[c]{@{}l@{}}0.928\\ $\pm$0.01\end{tabular}  & \multicolumn{2}{l|}{\cellcolor[HTML]{FFFFC7}\begin{tabular}[c]{@{}l@{}}2.704\\ $\pm$0.6\end{tabular}}      & \cellcolor[HTML]{FFCCC9}\begin{tabular}[c]{@{}l@{}}0.931\\ $\pm$0.009\end{tabular} & \multicolumn{2}{l|}{\cellcolor[HTML]{FFFFC7}\begin{tabular}[c]{@{}l@{}}2.794\\ $\pm$0.625\end{tabular}}    & \cellcolor[HTML]{FFCCC9}\begin{tabular}[c]{@{}l@{}}0.934\\ $\pm$0.008\end{tabular} & \multicolumn{2}{l|}{\cellcolor[HTML]{FFFFC7}\begin{tabular}[c]{@{}l@{}}2.377\\ $\pm$0.051\end{tabular}}          & \cellcolor[HTML]{FFCCC9}\begin{tabular}[c]{@{}l@{}}0.923\\ $\pm$0.012\end{tabular} & \multicolumn{2}{l|}{\cellcolor[HTML]{FFFFC7}\begin{tabular}[c]{@{}l@{}}2.648\\ $\pm$0.215\end{tabular}}    & \cellcolor[HTML]{FFCCC9}\begin{tabular}[c]{@{}l@{}}0.924\\ $\pm$0.002\end{tabular} & \multicolumn{2}{l|}{\cellcolor[HTML]{FFFFC7}\begin{tabular}[c]{@{}l@{}}2.787\\ $\pm$0.692\end{tabular}}    & \cellcolor[HTML]{FFCCC9}\begin{tabular}[c]{@{}l@{}}0.938\\ $\pm$0.008\end{tabular}          & \multicolumn{2}{l|}{\cellcolor[HTML]{FFFFC7}\begin{tabular}[c]{@{}l@{}}4.795\\ $\pm$2.619\end{tabular}}    \\ \hline
14                     & \begin{tabular}[c]{@{}c@{}}Left \\ Thalamus\end{tabular}             & \cellcolor[HTML]{FFCCC9}\begin{tabular}[c]{@{}l@{}}0.928\\ $\pm$0.01\end{tabular}  & \multicolumn{2}{l|}{\cellcolor[HTML]{FFFFC7}\begin{tabular}[c]{@{}l@{}}3.858\\ $\pm$1.212\end{tabular}}    & \cellcolor[HTML]{FFCCC9}\begin{tabular}[c]{@{}l@{}}0.931\\ $\pm$0.01\end{tabular}  & \multicolumn{2}{l|}{\cellcolor[HTML]{FFFFC7}\begin{tabular}[c]{@{}l@{}}4.527\\ $\pm$1.486\end{tabular}}    & \cellcolor[HTML]{FFCCC9}\begin{tabular}[c]{@{}l@{}}0.934\\ $\pm$0.009\end{tabular} & \multicolumn{2}{l|}{\cellcolor[HTML]{FFFFC7}\begin{tabular}[c]{@{}l@{}}2.544\\ $\pm$0.12\end{tabular}}           & \cellcolor[HTML]{FFCCC9}\begin{tabular}[c]{@{}l@{}}0.922\\ $\pm$0.013\end{tabular} & \multicolumn{2}{l|}{\cellcolor[HTML]{FFFFC7}\begin{tabular}[c]{@{}l@{}}2.896\\ $\pm$0.43\end{tabular}}     & \cellcolor[HTML]{FFCCC9}\begin{tabular}[c]{@{}l@{}}0.923\\ $\pm$0.002\end{tabular} & \multicolumn{2}{l|}{\cellcolor[HTML]{FFFFC7}\begin{tabular}[c]{@{}l@{}}3.219\\ $\pm$0.933\end{tabular}}    & \cellcolor[HTML]{FFCCC9}\begin{tabular}[c]{@{}l@{}}0.937\\ $\pm$0.01\end{tabular}           & \multicolumn{2}{l|}{\cellcolor[HTML]{FFFFC7}\begin{tabular}[c]{@{}l@{}}6.928\\ $\pm$2.766\end{tabular}}    \\ \hline
\multicolumn{1}{|l|}{} & \begin{tabular}[c]{@{}c@{}}Average\\  of \\ label 3-14\end{tabular}  & \cellcolor[HTML]{FFCCC9}\begin{tabular}[c]{@{}l@{}}0.888\\ $\pm$0.017\end{tabular} & \multicolumn{2}{l|}{\cellcolor[HTML]{FFFFC7}\begin{tabular}[c]{@{}l@{}}3.203\\ $\pm$0.368\end{tabular}}    & \cellcolor[HTML]{FFCCC9}\begin{tabular}[c]{@{}l@{}}0.894\\ $\pm$0.017\end{tabular} & \multicolumn{2}{l|}{\cellcolor[HTML]{FFFFC7}\begin{tabular}[c]{@{}l@{}}3.451\\ $\pm$0.406\end{tabular}}    & \cellcolor[HTML]{FFCCC9}\begin{tabular}[c]{@{}l@{}}0.897\\ $\pm$0.016\end{tabular} & \multicolumn{2}{l|}{\cellcolor[HTML]{FFFFC7}\textbf{\begin{tabular}[c]{@{}l@{}}2.793\\ $\pm$0.057\end{tabular}}} & \cellcolor[HTML]{FFCCC9}\begin{tabular}[c]{@{}l@{}}0.879\\ $\pm$0.026\end{tabular} & \multicolumn{2}{l|}{\cellcolor[HTML]{FFFFC7}\begin{tabular}[c]{@{}l@{}}3.126\\ $\pm$0.489\end{tabular}}    & \cellcolor[HTML]{FFCCC9}\begin{tabular}[c]{@{}l@{}}0.894\\ $\pm$0.003\end{tabular} & \multicolumn{2}{l|}{\cellcolor[HTML]{FFFFC7}\begin{tabular}[c]{@{}l@{}}3.005\\ $\pm$0.833\end{tabular}}    & \cellcolor[HTML]{FFCCC9}\textbf{\begin{tabular}[c]{@{}l@{}}0.902\\ $\pm$0.027\end{tabular}} & \multicolumn{2}{l|}{\cellcolor[HTML]{FFFFC7}\begin{tabular}[c]{@{}l@{}}5.123\\ $\pm$1.463\end{tabular}}    \\ \hline
                       & \begin{tabular}[c]{@{}c@{}}Full \\ Average\end{tabular}              & \cellcolor[HTML]{FFCCC9}\begin{tabular}[c]{@{}l@{}}0.865\\ $\pm$0.023\end{tabular} & \multicolumn{2}{l|}{\cellcolor[HTML]{FFFFC7}\begin{tabular}[c]{@{}l@{}}3.308\\ $\pm$0.332\end{tabular}}    & \cellcolor[HTML]{FFCCC9}\begin{tabular}[c]{@{}l@{}}0.872\\ $\pm$0.023\end{tabular} & \multicolumn{2}{l|}{\cellcolor[HTML]{FFFFC7}\begin{tabular}[c]{@{}l@{}}3.557\\ $\pm$0.442\end{tabular}}    & \cellcolor[HTML]{FFCCC9}\begin{tabular}[c]{@{}l@{}}0.877\\ $\pm$0.022\end{tabular} & \multicolumn{2}{l|}{\cellcolor[HTML]{FFFFC7}\textbf{\begin{tabular}[c]{@{}l@{}}2.94\\ $\pm$0.041\end{tabular}}}  & \cellcolor[HTML]{FFCCC9}\begin{tabular}[c]{@{}l@{}}0.856\\ $\pm$0.031\end{tabular} & \multicolumn{2}{l|}{\cellcolor[HTML]{FFFFC7}\begin{tabular}[c]{@{}l@{}}3.218\\ $\pm$0.467\end{tabular}}    & \cellcolor[HTML]{FFCCC9}\begin{tabular}[c]{@{}l@{}}0.867\\ $\pm$0.002\end{tabular} & \multicolumn{2}{l|}{\cellcolor[HTML]{FFFFC7}\begin{tabular}[c]{@{}l@{}}3.005\\ $\pm$0.734\end{tabular}}    & \cellcolor[HTML]{FFCCC9}\textbf{\begin{tabular}[c]{@{}l@{}}0.881\\ $\pm$0.065\end{tabular}} & \multicolumn{2}{l|}{\cellcolor[HTML]{FFFFC7}\begin{tabular}[c]{@{}l@{}}4.952\\ $\pm$1.23\end{tabular}}     \\ \hline
\end{tabular}
\end{minipage}
\end{sideways}
\caption{Performance of 3D Unet, Residual 3D Unet, Dense U-net and V-net with IBSD data in terms of dice coefficient and HD with respect to Ground truth. The values are in Mean$\pm$ Standard Deviation format}
\label{table:segperf}
\end{table}

\subsection*{Segmentation performance on other Datasets}
 It is well known that the performance of segmentation algorithms depend on many aspects of the dataset, such as number of images in the set, image quality and structural details. As a final experiment three models were trained and tested on two public datasets (IBSR and MICCAI 2012) and compared with that on IBSD.
 The obtained results are presented in Table \ref{table:comp}, with the best results for each method indicated in bold font. From this Table, it can be observed that training on IBSD yields best Dice consistently for all 3 DL methods (last 3 rows) and lowest HD value are obtained for 2 of 3 DL methods. Since IBSD has the maximum number of images, this result reaffirms the data dependent nature of DL and emphasises the need for a larger dataset for segmentation.  
\begin{table}[ht]
\begin{tabular}{|c|c|c|c|c|c|c|}
\hline
            & \multicolumn{2}{c|}{IBSR}                                                                 & \multicolumn{2}{c|}{MICCAI}                                                       & IBSD                      &                           \\ \hline
            & Dice                      & \begin{tabular}[c]{@{}c@{}}Hausdorff \\ Distance\end{tabular} & Dice              & \begin{tabular}[c]{@{}c@{}}Hausdorff \\ Distance\end{tabular} & Dice                      & HD                        \\ \hline

3D-Unet     & 0.842 $\pm$ 0.052         & 3.52 $\pm$ 0.86                                               & 0.840 $\pm$ 0.047 & 4.01 $\pm$ 1.87                                               & \textbf{0.865 $\pm$0.023} & \textbf{3.308 $\pm$0.332} \\ \hline
Dense U-net & 0.846 $\pm$ 0.092         & 3.41 $\pm$1.27                                                & 0.864 $\pm$ 0.055 & 3.24 $\pm$1.64                                                & \textbf{0.877 $\pm$0.022} & \textbf{2.94 $\pm$0.041}  \\ \hline
$\psi$-net  & 0.855 $\pm$ 0.074         & 3.02 $\pm$ 0.55                                               & 0.875 $\pm$ 0.056 & \textbf{2.76 $\pm$ 0.72}                                      & \textbf{0.881 $\pm$0.065} & 4.952 $\pm$1.23           \\ \hline
\end{tabular}
\caption{Free-surfer,FIRST, 3D U-net, Dense U-net and $\psi$-net
 performance compared with IBSR, MICCAI and IBSD data with average dice coefficient and HD for 14 sub-cortical structures.The values are in Mean$\pm$ Standard Deviation format}
\label{table:comp}
\end{table}


\section*{Code availability}

All the prepossessing steps and experiments were done with publicly available software. We have performed Bias Field correction and denoising with ANTS software commands \textit{N4BiasFiledCorrection} and \textit{DenoiseImage} repsectively from \url{http://stnava.github.io/ANTs/}. We have used \textit{BET} command from FSL available in \url{https://fsl.fmrib.ox.ac.uk/fsl/fslwiki/BET/} to perform skull stripping.We have used ITK-SNAP 3.8.0 to perform manual segmentation corrections which is available in \url{http://www.itksnap.org/pmwiki/pmwiki.php}.
The segmentation methods we have used in the experiment section are also publicly available. We have used Freesurfer from \url{https://surfer.nmr.mgh.harvard.edu/}, 3d unet from \url{ https://github.com/wolny/pytorch-3dunet}, residual 3d unet from \url{https://github.com/wolny/pytorch-3dunet}, v-net from \url{https://github.com/faustomilletari/VNet}, Mnet from \url{https://github.com/Prathyusha-Akundi/M-Net} and psi-net from \\  \url{https://github.com/lihaoliu-cambridge/psi-net}.

\bibliography{sample}

\section*{Acknowledgements}
We acknowledge IHub-Data, IIIT Hyderabad, India for financial assistance for publication.

\section*{Author contributions statement}

J.S. conceptualized the work, A.J. and M.V. equally  contributed the data analysis, R.M. collected the MRI scans while he was in IIIT-Hydearabad, S.K.(radiology expert ) and C.K.(Senior radiologist) done the manual segmentations. All authors reviewed the manuscript. 

\section*{Competing interests}
The authors declare no competing interests.

\end{document}